\documentclass[12pt]{article}

\begin{document}

\title{Scattering of neutral fermions by a pseudoscalar potential step in
two-dimensional space-time}
\date{}
\author{Antonio S. de Castro\thanks{%
E-mail address: castro@feg.unesp.br (A.S. de Castro)} \\
UNESP, Campus de Guaratinguet\'{a}\\
Departamento de F\'{i}sica e Qu\'{i}mica\\
Caixa Postal 205\\
12516-410 Guaratinguet\'{a} SP, Brasil}
\maketitle

\begin{abstract}
The problem of scattering of neutral fermions in two-dimensional space-time
is approached with a pseudoscalar potential step in the Dirac equation. Some
unexpected aspects of the solutions beyond the absence of Klein\'{}s paradox
are presented. An apparent paradox concerning the uncertainty principle is
solved by introducing the concept of effective Compton wavelength. Added
plausibility for the existence of bound-state solutions in a pseudoscalar
double-step potential found in a recent Letter is given.
\end{abstract}

\pagebreak

In a recent Letter the problem of confinement of neutral fermions by a
double-step potential with pseudoscalar coupling in 1+1 dimensions was
approached \cite{asc}. There it was found that this sort of potential
results in confinement in the event that the coupling is strong enough. The
one-dimensional Dirac equation developed in that Letter could also be
obtained from the four-dimensional one with the mixture of spherically
symmetric scalar, vector and anomalous magnetic interactions. If we had
limited the fermion to move in the $x$-direction ($p_{y}=p_{z}=0$) the
four-dimensional Dirac equation would decompose into two equivalent
two-dimensional equations with 2-component spinors and 2$\times $2 matrices 
\cite{str}. Then, there would result that the scalar and vector interactions
would preserve their Lorentz structures whereas the anomalous magnetic
interaction would turn out to be a pseudoscalar interaction. Furthermore, in
the (1+1)-world there is no angular momentum so that the spin is absent.
Therefore, the (1+1)-dimensional Dirac equation allow us to explore the
physical consequences of the negative energy states in a mathematically
simpler and more physically transparent way.

In the present Letter some aspects of the scattering of neutral fermions in
a pseudoscalar potential step in a (1+1) dimensional space-time are
analyzed. This essentially relativistic problem gives rise to some
unexpected results and make more credible the existence of the bound-state
solutions found in the former Letter \cite{asc}.

Let us begin by presenting the Dirac equation in 1+1 dimensions. In the
presence of a time-independent potential the (1+1)-dimensional
time-independent Dirac equation for a fermion of rest mass $m$ reads 
\begin{equation}
\mathcal{H}\Psi =E\Psi   \label{eq1}
\end{equation}

\begin{equation}
\mathcal{H}=c\alpha p+\beta mc^{2}+\mathcal{V}  \label{eq1a}
\end{equation}

\noindent where $E$ is the energy of the fermion, $c$ is the velocity of
light and $p$ is the momentum operator. $\alpha $ and $\beta $ are Hermitian
square matrices satisfying the relations $\alpha ^{2}=\beta ^{2}=1$, $%
\left\{ \alpha ,\beta \right\} =0$. From the last two relations it steams
that both $\alpha $ and $\beta $ are traceless and have eigenvalues equal to 
$-$1, so that one can conclude that $\alpha $ and $\beta $ are
even-dimensional matrices. One can choose the 2$\times $2 Pauli matrices
satisfying the same algebra as $\alpha $ and $\beta $, resulting in a
2-component spinor $\Psi $. The positive definite function $|\Psi |^{2}=\Psi
^{\dagger }\Psi $, satisfying a continuity equation, is interpreted as a
probability position density and its norm is a constant of motion. This
interpretation is completely satisfactory for single-particle states \cite
{tha}. We use $\alpha =\sigma _{1}$ and $\beta =\sigma _{3}$. For the
potential matrix we consider 
\begin{equation}
\mathcal{V}=1V_{t}+\beta V_{s}+\alpha V_{e}+\beta \gamma ^{5}V_{p}
\label{eq2}
\end{equation}

\noindent where $1$ stands for the 2$\times $2 identity matrix and $\beta
\gamma ^{5}=\sigma _{2}$. This is the most general combination of Lorentz
structures for the potential matrix because there are only four linearly
independent 2$\times $2 matrices. The subscripts for the terms of potential
denote their properties under a Lorentz transformation: $t$ and $e$ for the
time and space components of the 2-vector potential, $s$ and $p$ for the
scalar and pseudoscalar terms, respectively. It is worth to note that the
Dirac equation is invariant under $x\rightarrow -x$ if $V_{e}(x)$ and $%
V_{p}(x)$ change sign whereas $V_{t}(x)$ and $V_{s}(x)$ remain the same.
This is because the parity operator $P=\exp (i\theta )P_{0}\sigma _{3}$,
where $\theta $ is a constant phase and $P_{0}$ changes $x$ into $-x$,
changes sign of $\alpha $ and $\beta \gamma ^{5}$ but not of $1$ and $\beta $%
.

Defining the spinor $\psi $ as 
\begin{equation}
\psi =\exp \left( \frac{i}{\hbar }\Lambda \right) \Psi  \label{eq5}
\end{equation}

\noindent where 
\begin{equation}
\Lambda (x)=\int^{x}dx^{\prime }\frac{V_{e}(x^{\prime })}{c}  \label{eq6}
\end{equation}

\noindent the space component of the vector potential is gauged away

\begin{equation}
\left( p+\frac{V_{e}}{c}\right) \Psi =\exp \left( \frac{i}{\hbar }\Lambda
\right) p\psi  \label{eq7}
\end{equation}

\noindent so that the time-independent Dirac equation can be rewritten as
follows:

\begin{equation}
H\psi =E\psi  \label{eq7a}
\end{equation}

\begin{equation}
H=\sigma _{1}cp+\sigma _{2}V_{p}+\sigma _{3}\left( mc^{2}+V_{s}\right)
+1V_{t}  \label{eq8}
\end{equation}

\noindent showing that the space component of a vector potential only
contributes to change the spinors by a local phase factor.

Provided that the spinor is written in terms of the upper and the lower
components 
\begin{equation}
\psi =\left( 
\begin{array}{c}
\phi \\ 
\chi
\end{array}
\right)  \label{eq8a}
\end{equation}

\noindent the Dirac equation decomposes into :

\begin{eqnarray}
\left( V_{t}-E+V_{s}+mc^{2}\right) \phi (x) &=&i\hbar c\chi ^{\prime
}(x)+iV_{p}\chi (x)  \nonumber \\
&&  \label{eq8b} \\
\left( V_{t}-E-V_{s}-mc^{2}\right) \chi (x) &=&i\hbar c\phi ^{\prime
}(x)-iV_{p}\phi (x)  \nonumber
\end{eqnarray}

\noindent where the prime denotes differentiation with respect to $x$. In
terms of $\phi $ and $\chi $ the spinor is normalized as $\int_{-\infty
}^{+\infty }dx\left( |\phi |^{2}+|\chi |^{2}\right) =1$, so that $\phi $ and 
$\chi $ are square integrable functions. It is clear from the pair of
coupled first-order differential equations (\ref{eq8b}) that both $\phi (x)$
and $\chi (x)$ must be discontinuous wherever the potential undergoes an
infinite jump and have opposite parities if the Dirac equation is invariant
under $x\rightarrow -x$. In the nonrelativistic approximation (potential
energies small compared to the rest mass) Eq. (\ref{eq8b}) becomes

\begin{equation}
\chi =\frac{p}{2mc}\phi  \label{eq8c}
\end{equation}

\begin{equation}
\left( -\frac{\hbar ^{2}}{2m}\frac{d^{2}}{dx^{2}}+V_{t}+V_{s}\right) \phi
=\left( E-mc^{2}\right) \phi  \label{eq8d}
\end{equation}

\noindent Eq. (\ref{eq8c}) shows that $\chi $ if of order $v/c<<1$ relative
to $\phi $ and Eq. (\ref{eq8d}) shows that $\phi $ obeys the Schr\"{o}dinger
equation without any contribution from the pseudoscalar potential.

Now, let us choose the nonconserving-parity pseudoscalar potential step

\begin{equation}
V_{p}(x)=\left\{ 
\begin{array}{cc}
0, & x<0 \\ 
V_{0}, & x>0
\end{array}
\right.   \label{eq9}
\end{equation}

\noindent Then the solutions of the Dirac equation describing a fermion
coming in from the left can be written as

\begin{equation}
\psi (x)=\left\{ 
\begin{array}{cc}
A\left( 
\begin{array}{c}
1 \\ 
\mu _{1}
\end{array}
\right) e^{+ik_{1}x}+A^{^{\prime }}\left( 
\begin{array}{c}
1 \\ 
-\mu _{1}
\end{array}
\right) e^{-ik_{1}x}, & x<0 \\ 
&  \\ 
B\left( 
\begin{array}{c}
1 \\ 
i\mu _{2}
\end{array}
\right) e^{+ik_{2}x}, & x>0
\end{array}
\right.  \label{eq10}
\end{equation}

\noindent where

\begin{eqnarray}
\left( \hbar ck_{1}\right) ^{2} &=&E^{2}-m^{2}c^{4}  \nonumber \\
&&  \nonumber \\
\left( \hbar ck_{2}\right) ^{2} &=&E^{2}-m^{2}c^{4}-V_{0}^{2}  \nonumber \\
&&  \label{eq11} \\
\mu _{1} &=&\frac{\hbar ck_{1}}{E+mc^{2}}  \nonumber \\
&&  \nonumber \\
\mu _{2} &=&\frac{-i\hbar ck_{2}+V_{0}}{E+mc^{2}}  \nonumber
\end{eqnarray}

\noindent

\noindent The second line of (\ref{eq11}) can be written in the more
suggestive form

\begin{equation}
\left( \hbar ck_{2}\right) ^{2}=E^{2}-m^{*2}c^{4}  \label{eq11a}
\end{equation}

\noindent where the effective mass

\begin{equation}
m^{*}=\sqrt{m^{2}+V_{0}^{2}/c^{2}}  \label{eq11b}
\end{equation}

\noindent is independent of the sign of $V_{0}$.

The spinor for $x<0$ describes the incident wave moving to the right ($k_{1}$
is a real number) and the reflected wave moving to the left. The spinor for $%
x>0$ describes a transmitted wave moving to the right if $k_{2}$ is a real
number, otherwise it describes an attenuated exponentially wave beyond the
potential boundary

\smallskip

The continuity of the spinor \noindent \ at the potential boundary yields
the relative amplitudes

\begin{eqnarray}
r &\equiv &\frac{A^{^{\prime }}}{A}=\frac{\hbar c\left( k_{1}-k_{2}\right)
-iV_{0}}{\hbar c\left( k_{1}+k_{2}\right) +iV_{0}}  \nonumber \\
&&  \label{eq12} \\
t &\equiv &\frac{B}{A}=\frac{2\hbar ck_{1}}{\hbar c\left( k_{1}+k_{2}\right)
+iV_{0}}  \nonumber
\end{eqnarray}

\noindent In order to determinate the reflection and transmission
coefficients we use the probability current density $J=c\psi ^{\dagger
}\alpha \psi $, where $\psi ^{\dagger }$ is the Hermitian conjugate spinor.
The probability current densities in the two different regions of the $x$%
-axis are

\begin{eqnarray}
J(x &<&0)=\frac{2\hbar c^{2}k_{1}}{E+mc^{2}}\left( |A|^{2}-|A^{^{\prime
}}|^{2}\right)  \nonumber \\
&&  \label{eq13} \\
J(x &>&0)=\frac{2\hbar c^{2}\Re(k_{2})}{E+mc^{2}}|B|^{2}  \nonumber
\end{eqnarray}

\noindent so that the $x$-independent current density allows us to define
the reflection and transmission coefficients as

\begin{eqnarray}
R &\equiv &|r|^{2}=\frac{\hbar ^{2}c^{2}\left[ k_{1}^{2}+|k_{2}|^{2}-2k_{1}%
\Re(k_{2})\right] +V_{0}^{2}+2\hbar cV_{0}\Im(k_{2})}{\hbar
^{2}c^{2}\left[ k_{1}^{2}+|k_{2}|^{2}+2k_{1}\Re(k_{2})\right]
+V_{0}^{2}+2\hbar cV_{0}\Im(k_{2})}  \nonumber \\
&&  \label{eq14} \\
T &\equiv &\frac{\Re(k_{2})}{k_{1}}|t|^{2}=\frac{4\hbar ^{2}c^{2}k_{1}%
\Re(k_{2})}{\hbar ^{2}c^{2}\left[ k_{1}^{2}+|k_{2}|^{2}+2k_{1}\Re%
(k_{2})\right] +V_{0}^{2}+2\hbar cV_{0}\Im(k_{2})}  \nonumber
\end{eqnarray}

\noindent One can note that $R+T=1$ and $R\ $is always less than one, so
that Klein\'{}s paradox is lacking.

There are two clearly marked behaviors for the spinors. If $%
E^{2}<V_{0}^{2}+m^{2}c^{4}$ then $k_{2}$ becomes purely imaginary and $\psi
(x>0)$ describes an evanescent wave ($E<m^{*}c^{2}$) with $R=1$ and $T=0$.
Even though $\psi (x>0)$ exists there is no probability current density
through the potential frontier. Conversely, if $E^{2}>V_{0}^{2}+m^{2}c^{4}$
then $k_{2}$ becomes a real number  so that $\psi (x>0)$ in fact describes a
transmitted wave ($E>m^{*}c^{2}$), with $R<1$ and $T\neq 0$.

Now let us move on to analyze the condition $E^{2}>V_{0}^{2}+m^{2}c^{4}$
which ensures that the spinor is an oscillatory wave ($k_{2}$ is a real
number) in the region $x>0$. The resulting condition turns into $-\sqrt{%
E^{2}-m^{2}c^{4}}<V_{0}<\sqrt{E^{2}-m^{2}c^{4}}$ ($m<m^{*}<E/c^{2}$). This
resembles the nonrelativistic result for a ascending step. On the other
side, a deep enough step ($V_{0}<-\sqrt{E^{2}-m^{2}c^{4}}$) leads to total
reflection, in contrast to nonrelativistic quantum mechanics that furnishes $%
R=1$ only in the $V_{0}\rightarrow -\infty $ case.

The effective mass given by (\ref{eq11b}) clearly indicates that this kind
of potential couples to the mass of the fermion and consequently it couples
to the positive-energy component of the spinor in the same way it couples to
the negative-energy component. The energy spectrum for a free fermion ranges
continuously from $-mc^{2}$ to $-\infty $ as well as from $mc^{2}$ to $%
\infty $ and the potential step contributes to enhance the energy gap
bringing into being at $x=0$ an ascending step for the positive-energy
solutions and a descending step for the negative-energy solutions. The
spectral gap always stretches whichever the sign of $V_{0}$. This is so
because the effective mass depends on $V_{0}^{2}$. In other words, the
potential step at $x=0$ is repulsive for the particles as well as for the
antiparticles. In this way one can see that there is no room for transitions
from positive- to negative-energy solutions. It all means that Klein\'{}s
paradox does not come to the scenario.

Another interesting way of visualizing the absence of Klein\'{}s paradox is
to consider the uncertainty in the position beyond the potential boundary in
the case $E^{2}<V_{0}^{2}+m^{2}c^{4}$ ($E<m^{*}c^{2}$). In this circumstance
the probability density for $x>0$ is given by

\begin{equation}
\rho (x>0)=|B|^{2}e^{-2|k_{2}|x}\left[ 1+\left( \frac{\hbar c|k_{2}|+V_{0}}{%
E+mc^{2}}\right) ^{2}\right]  \label{eq15}
\end{equation}

\noindent so that

\begin{equation}
\Delta x=\frac{\hbar c}{2\sqrt{m^{2}c^{4}+V_{0}^{2}-E^{2}}}  \label{eq16}
\end{equation}

\noindent From this last result one can see that the penetration of the
fermion into the region $x>0$ will shrink without limit as $|V_{0}|$
increases and there is no atmosphere for the production of
particle-antiparticle pairs at the potential frontier. It means that the
localization of a fermion under the influence of this sort of pseudoscalar
potential does not require any minimum value in order to ensure the
single-particle interpretation of the Dirac equation. At first glance it
seems that the uncertainty principle dies away provided such a principle
implies that it is impossible to localize a particle into a region of space
less than half of its Compton wavelength (see, \textit{e.g.}, Ref \cite{str}%
). This apparent contradiction can be remedied by writing (\ref{eq16}) into
the form

\begin{equation}
\Delta x=\frac{\hbar c}{2\sqrt{m^{*2}c^{4}-E^{2}}}  \label{eq17}
\end{equation}

\noindent Thus, it can be  immediately seen \noindent that the minimum
uncertainty is

\begin{equation}
\Delta x_{\min }=\frac{\hbar }{2m^{*}c}  \label{eq18}
\end{equation}

\noindent in the limit as $|V_{0}|\rightarrow \infty $, and again the sign
of $V_{0}$ plays no part. Therefore, for obtaining a result consistent with
the uncertainty principle it is a must to use the effective Compton
wavelength.

In addition to their intrinsic importance, the above conclusions for the
scattering of a neutral fermion by a pseudoscalar single-step potential
enhance the plausibility arguments for the confinement of neutral fermions
by a pseudoscalar double-step potential presented in the former Letter \cite
{asc}.

We have made the analysis of the scattering of neutral fermions by a
pseudoscalar potential step in the (1+1)-dimensional Dirac equation. The
results obtained in the (1+1) world are also supposed to be qualitatively
true in the more realistic 3+1 world. Although the idea and calculations
presented in this Letter are simple the results and conclusions seem to be
new.

\bigskip

\smallskip

\noindent \textbf{Acknowledgments}

This work was supported in part through funds provided by CNPq and FAPESP.

\end{document}